\documentclass[12pt]{article}
\usepackage[main=english]{babel}
\usepackage[T1]{fontenc}
\usepackage{graphicx} 
\usepackage{amsmath}
\usepackage{amsthm}
\usepackage{amsfonts}
\usepackage{amssymb}
\usepackage{derivative}
\usepackage[toc,page]{appendix}
\usepackage[sorting=none, maxnames=70]{biblatex}
\usepackage{csquotes}
\usepackage[affil-it]{authblk}
\usepackage{indentfirst}
\usepackage{enumitem} 
\usepackage{xcolor}

\title{Herglotz's formalism, Eisenhart lift and Killing vectors}
\author{Krystian Bartczak%
    \thanks{krystian.bartczak@edu.uni.lodz.pl}\;}
\author{Piotr Kosiński%
    \thanks{piotr.kosinski@uni.lodz.pl}}

\affil{Faculty of Physics and Applied Informatics, University of Lodz, Poland}
\date{}

\renewbibmacro{in:}{}
\DeclareFieldFormat{pages}{#1}

\makeatletter
\def\@maketitle{%
  \newpage
  \null
  \vskip 2em%
  \begin{center}%
  \let\footnote\thanks
    {\Large\@title\par}%
    \vskip 1.5em%
    {\normalsize
      \lineskip.5em%
      \begin{tabular}[t]{c}%
        \@author
      \end{tabular}\par}%
    \vskip 1em%
    {\normalsize \@date}%
  \end{center}%
  \par
  \vskip 1.5em}
\makeatother

\renewcommand{\appendix}{
    \setcounter{section}{0}
    \setcounter{subsection}{0}
    \renewcommand{\thesection}{\Alph{section}}

    \counterwithin{figure}{section}
    \counterwithin{table}{section}
    \phantomsection
    \addcontentsline{toc}{part}{\appendixname} 

    \clearpage
    \thispagestyle{plain}
    {\centering
     \normalfont\Huge
     \textbf{\appendixname}\par}

}

\DefineBibliographyExtras{english}{
  
}

\begin{document}
    \maketitle

    \begin{abstract}

The Eisenhart lift is extended to the case of dynamics described by action-dependent Lagrangians.
The resulting Brinkmann metric depends on all coordinates.
It is shown that the symmetries of the initial dynamics result in the existence of
(conformal) Killing vectors of the Brinkmann metric.
An example is given of equivalent time- and action-dependent descriptions which result in
conformally equivalent metrics. It is also shown how the scaling invariance fits naturally into this scheme.
 
\end{abstract}

    \newpage

    \tableofcontents

  \begin{section}{Introduction}

A geometric approach to classical dynamics, currently known as Eisenhart lift,
was originally proposed by Eisenhart in 1928~\cite{Eisenhart} and then rediscovered by
a number of authors~\cite{Gomis,Bargmann,Balachandran,Celestial,Fordy} (for a review see~\cite{HiddenSym}).\\
Eisenhart lift relates the Newtonian dynamics in $n$ dimensions to the geodesic motion in $(n+2)$ dimensional (pseudo-)Riemannian manifold
equipped with the Brinkmann metric~\cite{Brinkmann,Celestial}. The two additional variables are identified with time and action.
Originally, the Eisenhart lift concerned conservative systems; however, the generalization
to the time-dependent Lagrangians is quite straightforward~\cite{TimeDependent}.
In the latter case the components of the Brinkmann metric tensor depend, in general, on $(n+1)$ coordinates.
The most general case of the metric depending on $(n+2)$ coordinates is more problematic.
This is because the last coordinate is identified, as mentioned above, with the action equal to the time integral of the Lagrange function.
However, there exists a generalization of the Lagrangian (and Hamiltonian) formalism which admits the Lagrange functions depending on additional variable which can
be viewed as a generalization of the action. This formalism is called Herglotz's variational principle~\cite{Herglotz,HerlotzLect,ActionPrinciple,HergVarProb}.
It appears that the Herglotz dynamical systems occur naturally when the Lagrangian dynamics is reduced to lower-dimensional configuration space.
Namely, the Lagrangian dynamics defined by the Lagrangian homogeneous of degree two in velocities, when reduced to the configuration space
one dimension lower, is described, in general, by action-dependent Lagrangians~\cite{Piwnik}.
Guided by this result we consider here the Eisenhart lift procedure for the metric depending on all coordinates.
We show that the general Brinkmann metric in $(n+2)$-dimensional space-time can be obtained by lifting the dynamics in $n$ dimensions provided the latter is described by an action-dependent Lagrangian.
Inherent in Eisenhart lift relation between symmetries, conservation laws and (conformal) Killing vectors is extended to the most general Brinkmann metric.
It is also shown in a simple example how the equivalence of the descriptions of dissipative dynamics in terms of time-dependent and action-dependent Lagrangians
is reflected in conformal equivalence of the relevant Brinkmann metrics.

One should admit that the Herglotz formalism is not particularly popular amongst physicists. However, it is closely related to the theory of contact transformations, a natural generalization of symplectic ones.
Therefore, it finds applications in thermodynamics~\cite{ContactFriction, Mrugala, Meso,HamiltonianExtension}, electromagnetic dissipative systems~\cite{EnM}, 
hydrodynamics~\cite{Hydro}, generalization of Einstein's equations in GR~\cite{ActionHilbert,GeneralizedGravField}, optimal control theory~\cite{OptimalControl}, dark energy~\cite{NoDarkEnergy}, cosmology~\cite{Sloan} or light propagation in general gravitational field~\cite{Piwnik}.
It should be also mentioned that making the relevant metric action-dependent destroys the property that it provides the solution to vacuum
Einstein's equations. However, the latter property is by no means necessary in the context of Eisenhart lift the essence of which is the geometrization of dynamics.

The paper is organized as follows. In Sec.\;\ref{sec:reduction} we briefly review the general result concerning the ``dimensional reduction'' of dynamics described by the Lagrangians homogeneous in velocities.
Then, in Sec.\;\ref{sec:lift}, the standard Eisenhart lift is described. Sec.\;\ref{sec:general} is devoted to the extension of the Eisenhart formalism to the case of general Brinkmann metrics.
It is shown that the latter emerges from lifting the dynamics described by Herglotz's variational principle.
Sec.\;\ref{sec:transf} contains a small remark concerning the role of point transformations in Brinkmann manifolds.
It is shown further in Sec.\;\ref{sec:sym} that the relation between symmetry transformations, Noether's theorem and Killing vectors, established in the framework
of standard Eisenhart lift, extends naturally to the general case described here.
In Sec.\;\ref{sec:timeaction} a simple example is given how the equivalence between time-dependent and action-dependent descriptions manifests itself on the level of Brinkmann metrics.
In Sec.\;\ref{sec:scaling} we consider the scaling symmetry showing that the Herglotz formalism provides a simple and natural framework for this kind of symmetry.
The Eisenhart lift leads in this case to conformal transformations of Brinkmann metrics.
Finally, Sec.\;\ref{sec:summary} is devoted to some summary.
In Appendix the Herglotz formalism is briefly described and some remarks concerning the reparametrization invariance are provided.

\end{section}
    \newpage

\begin{section}{Dimensional reduction of dynamics}\label{sec:reduction}
    
Eisenhart lift allows us to describe the dynamics of $n$ degrees of freedom in terms of geodesic motion in an
appropriately chosen (pseudo-)Riemannian metric. In the original version, one starts with Lagrangian
(or Hamiltonian) dynamics; the lifted dynamics is also described by the Lagrangian which is homogeneous
of degree two in generalized velocities. However, in general it cannot be taken for granted that the reduction
of geodesic motion to lower dimension results in Lagrangian dynamics. Still, the reduction procedure makes sense
provided one admits a more general notion of Lagrangian dynamics; namely, one can consider Lagrangians depending on
the action variable which results in generalization of variational principle known as Herglotz's variational principle~\cite{Herglotz,HerlotzLect,ActionPrinciple,HergVarProb}
(see Appendix for a brief description).
Once such a generalization is admitted the following general theorem can be proven~\cite{Piwnik}.
Consider a system with $(n+1)$ degrees of freedom $q^\mu$, $\mu=0,1,\dots,n$, described by the Lagrangian $L=L(\underline{q},\underline{\dot{q}})$;
here dot denotes the differentiation with respect to some evolution parameter $\sigma$. One assumes that: 
\begin{enumerate}[label=(\roman*)]
\item $L$ is nondegenerate, $\det{\left(   \pdv{L}{    \dot{q}^\mu,    \dot{q}^\nu}    \right)}\neq 0$;
\item $L$ is a homogeneous function of generalized velocities $\dot{q}^\mu$ of degree~$N\neq~1$;
\item $L$ does not depend explicitly on the evolution parameter $\sigma$.
\end{enumerate}
It follows then that the value of the Lagrangian is a constant of motion.
Let us fix the value $C$ of $L$ and consider the subset of trajectories obeying
        \begin{equation}\label{subset}
            L(\underline{q},\underline{\dot{q}})=C
        \end{equation}

Equation~\eqref{subset} can be solved, at least locally, with respect to $\dot{q}^0$:
        \begin{equation}\label{q0}
            \odv{q^0}{\sigma}=\mathcal{L}\left(q^0,q^1,\dots,q^n,\dot{q}^1,\dots,\dot{q}^n;C\right)
        \end{equation}
{Actually, eq.~\eqref{subset} admits, in general, several local inverse functions.
However, in the case we are here interested in (Brinkmann metric), eq.~\eqref{subset} possesses a unique solution.
}

The projection of initial dynamics on the generalized coordinates $q^1,\dots,q^n$ is described by the Herglotz equations:
        \begin{align}\label{hereq}
            \pdv{\mathcal{L}}{q^i}-\odv*{\left(\pdv{\mathcal{L}}{\dot{q}^i}\right)}{\sigma}+\pdv{\mathcal{L}}{q^0}\pdv{\mathcal{L}}{\dot{q}^i}=0,&\quad i=1,\dots,n
        \end{align}
together with eq.~\eqref{q0}.
We conclude that the reduced configuration space is spanned by $q^1,\dots,q^n$ while $q^0$ plays the role of action.

The particularly interesting case concerns the Lagrangians $L$ being the polynomials of degree 2 ($N=2$) in velocities:
        \begin{equation}
            L=\frac{1}{2} g_{\mu \nu}(\underline{q})\dot{q}^\mu \dot{q}^\nu
        \end{equation}

Assume that $g_{\mu\nu}(\underline{q})$ has the Lorentz signature. Consider the set of trajectories corresponding to $L=0$.
Then $\mathcal{L}$, defined by eq.~\eqref{q0}, is homogeneous of first degree and defines reparametrization invariant dynamics.
Therefore, one can choose the evolution parameter $\sigma$ at will;
for example, one can set $\sigma=q^n$, thereby reducing the number of degrees of freedom by two.

{Although the case $N=2$ seems to be the most interesting one, we described here the general case $N \neq 1$.
This is because for any $N\neq 1$ the proof goes along the same lines.
Moreover, the case $N=0$ covers the so-called super-Lagrangian formulation of Herglotz dynamics \cite{HergVarProb}.}

Finally, let us note that $\mathcal{L}$ defines standard Lagrangian mechanics if $L$ does not depend on $q^0$. 
    
{In this section we considered the problem of dimensional reduction, transforming the initial dynamics 
(which, in the most interesting case, describes geodesic motion) into lower dimensional,
in general non-geodesic system.
The Eisenhart lift, described in the next sections, goes in the opposite direction.
However, in the case of Brinkmann metric, eq.~\eqref{subset} has a unique solution.
Therefore, as we shall see, dimensional reduction and Eisenhart lift are mutually inverse operations.}
\end{section}
\begin{section}{Eisenhart lift}\label{sec:lift}

The standard Eisenhart lift deals with the $(n+2)$-dimensional manifold parametrized by the coordinates $x^1,\dots,x^n,u,w$, equipped with
the so-called Brinkmann metric~\cite{Brinkmann,Celestial}:
        \begin{equation}\label{brinkmann}
            \odif{s}^2=g_{\mu\nu}(\underline{x})\odif{x}^\mu \odif{x}^\nu=h_{ij}(\underline{x})\odif{x}^i \odif{x}^j+2A_i(\underline{x})\odif{x}^i \odif{u} - 2V(\underline{x})\odif{u}^2 -2\odif{u}\odif{w}
        \end{equation}

Assuming $\sigma$ to be an affine parameter one concludes that the geodesics are described by the Lagrangian:
        \begin{equation}\label{geolagrangian}
            L=\frac{1}{2}h_{ij}(\underline{x})\dot{x}^i \dot{x}^j+A_i(\underline{x})\dot{x}^i\dot{u}-V(\underline{x})\dot{u}^2 - \dot{u}\dot{w}
        \end{equation}

Lagrange equations resulting from~\eqref{geolagrangian} are equivalent to the geodesic equations for Brinkmann metric:
    
        \begin{equation}\label{lagrangiangeo}
            \ddot{x}^i+\Gamma^i_{jk}\dot{x}^j \dot{x}^k + h^{ik}F_{jk} \dot{x}^j \dot{u} + h^{ik} \pdif{k}{V}\dot{u}^2=0
        \end{equation}

        \begin{equation}\label{uddot}
            \ddot{u}=0
        \end{equation}

        \begin{equation}
            \ddot{w}+\Bigl(A_k\Gamma^k_{ij}-\frac{1}{2}\left(\pdif{i}{A_j}+\pdif{j}{A_i}\right)\Bigr)\dot{x}^i \dot{x}^j + \Bigl(A_j h^{jk}F_{ik}+2\pdif{i}{V}\Bigr)\dot{x}^i \dot{u} + A_j h^{ij} \pdif{i}{V} \dot{u}^2=0
        \end{equation}
here $\Gamma^i_{jk}$ are Christoffel symbols for the $h_{ij}$ metric, $h^{ij}$ is the inverse of $h_{ij}$ and $F_{ij}=\pdif{i}{A_j}-\pdif{j}{A_i}$.

Eq.~\eqref{uddot} implies that $u$ may be viewed as the affine parameter (the latter is defined up to an affine transformation).
Choosing additionally the trajectories with $L=0$ one concludes that
        \begin{equation}\label{wdot}
          \dot{w}=\dot{u}\left(\frac{1}{2}h_{ij} \frac{\dot{x}^i \dot{x}^j}{\dot{u}^2}+A_i \frac{\dot{x}^i}{\dot{u}}-V\right)
        \end{equation}    
Eqs.~\eqref{lagrangiangeo} and~\eqref{wdot} imply finally
        \begin{equation}\label{red}
            \odv[order=2]{x^i}{u}+\Gamma^i_{jk}\odv{x^j}{u}\odv{x^k}{u}+h^{ik}F_{jk}\odv{x^j}{u}+h^{ik}\pdif{k}{V}=0\
        \end{equation}

        \begin{equation}\label{newlag}
            \odv{w}{u}=\frac{1}{2}h_{ij}\odv{x^i}{u}\odv{x^j}{u}+A_i\odv{x^i}{u}-V=\mathcal{L}\left(\underline{x},\underline{\odv{x}{u}}\right)
        \end{equation}

By noting that~\eqref{red} are the Lagrange equations resulting from $\mathcal{L}\left(\underline{x},\underline{\odv{x}{u}}\right)$ we conclude that $u$ may be identified with time
while $w$ with the action for a particle moving in the curved space (with the metrics $h_{ij}$) under the influence of the scalar and vector potentials $V$ and $A_i$, respectively.

It is not difficult to generalize the above reasoning to the case of $u$-dependent (i.e. $t$-dependent) metric and potentials.

{Let us note that the Eisenhart lift fits precisely into the algorithm described in Sec.~\ref{sec:reduction}.
Here $w$ plays the role of $q^0$ while $C$ (eq.~\eqref{subset}) equals $0$.
As a result of reparametrization invariance in this case, $u$ can be chosen as evolution parameter.
Moreover, since the metric does not depend on $w$, one starts with an ordinary Lagrangian.}
\end{section}
 \begin{section}{Eisenhart lift - the general case}\label{sec:general}

Consider now the general Brinkmann metric depending on all coordinates $x^\mu=(x^k,u,w)$:
        \begin{equation}\label{metricgeneral}
            \odif{s}^2=h_{ij}(\underline{x},u,w)\odif{x^i,x^j}+2A_i(\underline{x},u,w)\odif{x^i,u}-2V(\underline{x},u,w)\odif{u}^2-2\odif{u,w}
        \end{equation}
or
        \begin{equation}
            L=\frac{1}{2}h_{ij}(\underline{x},u,w)\dot{x}^i \dot{x}^j + A_i(\underline{x},u,w)\dot{x}^i \dot{u} - V(\underline{x},u,w)\dot{u}^2-\dot{u}\dot{w}
        \end{equation}
$L$ is again an ordinary Lagrangian describing dynamics of $(n+2)$ degrees of freedom. We shall show that, in the spirit of the result described in~Sec.\;\ref{sec:reduction},
$L$ can be viewed as describing the Eisenhart-like lift of $n$-dimensional Herglotz dynamics. The relevant Lagrange equations (geodesic equations for the metric~\eqref{metricgeneral}) read:
        \begin{equation}\label{geogeneral}
h_{kj}\ddot{x}^j+\Gamma_{kij}\dot{x}^i\dot{x}^j+\dot{u}^2\pdif{k}{V}+A_k \ddot{u}+F_{ik}\dot{x}^i\dot{u}+\pdif{u}{h_{jk}}\dot{x}^j\dot{u}+\pdif{u}{A_k}\dot{u}^2+\pdif{w}{h_{jk}}\dot{x}^j\dot{w}+\pdif{w}{A_k}\dot{u}\dot{w}=0    
        \end{equation}

        \begin{equation}\label{ugeneral}
                \ddot{u}+\frac{1}{2}\pdif{w}{h_{ij}}\dot{x}^i\dot{x}^j+\pdif{w}{A_i}\dot{x}^i\dot{u}-\pdif{w}{V}\dot{u}^2=0
        \end{equation}

        \begin{equation}\label{wgeneral}
            \ddot{w}+\frac{1}{2}\pdif{u}{h_{ij}}\dot{x}^i\dot{x}^j+\pdif{u}{A_i}\dot{x}^i\dot{u}-\pdif{u}{V}\dot{u}^2-\odv*{\left(A_i\dot{x}^i-2V\dot{u}\right)}{\sigma}=0
        \end{equation}
Again, let us impose the condition $L=0$; \\
this yields
        \begin{equation}\label{laginvariant}
            \dot{w}=\dot{u}\left(\frac{1}{2}h_{ij}\frac{\dot{x}^i\dot{x}^j}{\dot{u}^2}+A_i\frac{\dot{x}^i}{\dot{u}}-V\right)
        \end{equation}
Eq.~\eqref{laginvariant} is reparametrization invariant.
It can be rewritten as
        \begin{equation}\label{afterreduction}
            \odv{w}{u}=\frac{1}{2}h_{ij}\odv{x^i}{u}\odv{x^j}{u}+A_i\odv{x^i}{u}-V\equiv \mathcal{L}\left(\underline{x},\underline{\odv{x}{u}},u,w\right)
        \end{equation}
Eq.~\eqref{afterreduction} suggests that, in the framework of Herglotz formalism, $u$ can be identified with time while $w$ -- with action.
In order to show that this interpretation is correct we consider eq.~\eqref{geogeneral} and~\eqref{ugeneral} (eq.~\eqref{wgeneral} is redundant once we imposed the constraint~\eqref{laginvariant}).
\newpage
\noindent Dividing them by $\dot{u}^2$ one finds:
        \begin{equation}\label{uddotu2}
            \frac{\ddot{u}}{\dot{u}^2}=-\pdif{w}{\mathcal{L}}
        \end{equation}

        \begin{equation}\label{beforeapply}
            h_{kj}\frac{\ddot{x}^j}{\dot{u}^2}+\Gamma_{kij}\frac{\dot{x}^i\dot{x}^j}{\dot{u}^2}+\pdif{k}{V}+A_k\frac{\ddot{u}}{\dot{u}^2}+F_{ik}\frac{\dot{x}^i}{\dot{u}}+\pdif{u}{\left(h_{jk}\frac{\dot{x}^j}{\dot{u}}+A_k\right)}+\pdif{w}{\left(h_{jk}\frac{\dot{x}^j}{\dot{u}^2}+\frac{A_k}{\dot{u}}\right)}\dot{w}=0
        \end{equation}
Moreover,
        \begin{equation}\label{step1}
            \odv{}{\sigma}=\dot{u}\odv{}{u}
        \end{equation}

        \begin{equation}\label{step2}
            \frac{1}{\dot{u}^2}\odv[order=2]{}{\sigma}=\frac{\ddot{u}}{\dot{u}^2}\odv{}{u}+\odv[order=2]{}{u}=-\pdif{w}{\mathcal{L}}\odv{}{u}+\odv[order=2]{}{u}
        \end{equation}
Using~\eqref{afterreduction},~\eqref{step1} and~\eqref{step2} one can rewrite eq.~\eqref{beforeapply} in terms of $u$-derivatives:
        \begin{align}\label{inu}
            &h_{kj}\odv[order=2]{x^j}{u}+\Gamma_{kij}\odv{x^i}{u}\odv{x^j}{u}+\nonumber\\
            +\;&\pdif{k}{V}+F_{ik}\odv{x^i}{u}+\pdif{u}{\left(h_{ik}\odv{x^i}{u}+A_k\right)}+\nonumber\\
            +\;&\pdif{w}{\left(h_{ik}\odv{x^i}{u}+A_k\right)}\mathcal{L}-\left(h_{ik}\odv{x^i}{u}+A_k\right)\pdif{w}{\mathcal{L}}=0
        \end{align}
It is now easy to check (again using~\eqref{afterreduction}) that eq.~\eqref{inu} is the Herglotz equations for $\mathcal{L}$:
        \begin{equation}
            \pdv{\mathcal{L}}{x^k}-\odv*{\left(\pdv{\mathcal{L}}{\!\left(\odv{x^k}{u}\right)}\right)}{\sigma}+\pdv{\mathcal{L}}{w}\pdv{\mathcal{L}}{\!\left(\odv{x^k}{u}\right)}=0
        \end{equation}
which, together with~\eqref{afterreduction}, forms the basis of Herglotz variational principle.

Let us stress that the dynamics resulting from the constraint $L=0$ is reparametrization invariant. Therefore, we can keep $\sigma$ as an evolution parameter
when deriving the reduced equations of motion and put $\sigma=u$ at the very end (this is briefly sketched in Appendix).
Concluding, we have shown that the Eisenhart lift can be applied to the quadratic, action- and time-dependent Lagrangian. The relevant $(n+2)$-dimensional metric takes the Brinkmann form with the metric tensor
components depending also on $w$.

{Again we see, in the general case, that the Eisenhart lift is closely related (as an inverse operation) to the reduction procedure described in Section~\ref{sec:reduction}.}
\end{section}   
\begin{section}{A note on point transformations}\label{sec:transf}

The Eisenhart lift allows us to relate the point transformations of initial dynamics to coordinate transformations in $(n+2)$-dimensional manifold equipped with Brinkmann metric.
Consider an arbitrary coordinate transformation in the latter
        \begin{equation}\label{transformation}
            {x^\prime}^\mu={x^\prime}^\mu(x^\nu),\quad \mu,\nu=0,1,\dots,n,n+1
        \end{equation}
with $x^0=w$, $x^{n+1}=u$.
In general,~\eqref{transformation}~does not preserve the Brinkmann form~\eqref{metricgeneral} of the metric.
This is because in the Herglotz formalism the transformation rule for the Lagrangian takes, in general, the form of an implicit relation\footnote{We thank Tomasz Ruciński for the enlightening discussion on this point.}.
As a result the new Lagrangian is no longer quadratic in velocities (this is also the case in the standard Lagrangian theory if the new time variable depends on old coordinates).
However, in view of the theorem described in~Sec.\;\ref{sec:reduction}, eq.~\eqref{transformation} should imply the correct rule for the transformation of action-dependent Lagrangian.
Eq.~\eqref{transformation} yields
        \begin{equation}
            \odv{{x^\prime}^0}{\sigma}=\pdv{{x^\prime}^0}{x^0}\odv{x^0}{\sigma}+\sum_{k=1}^n \pdv{{x^\prime}^0}{x^k}\odv{x^k}{\sigma}+\pdv{{x^\prime}^0}{x^{n+1}}\odv{x^{n+1}}{\sigma}
        \end{equation}
Multiplying by ${\left(\odv{x^{n+1}}{\sigma}\right)}^{-1}$ one finds
        \begin{equation}\label{transformationlaw1}
            {\left(\odv{x^{n+1}}{\sigma}\right)}^{-1}\odv{{x^\prime}^0}{\sigma}=\pdv{{x^\prime}^0}{x^0}\odv{x^0}{\sigma}{\left(\odv{x^{n+1}}{\sigma}\right)}^{-1}+\sum_{k=1}^n \pdv{{x^\prime}^0}{x^k}\odv{x^k}{\sigma}{\left(\odv{x^{n+1}}{\sigma}\right)}^{-1}+\pdv{{x^\prime}^0}{x^{n+1}}
        \end{equation}
or
        \begin{equation}\label{transformationlaw2}
            \odv{{x^\prime}^{n+1}}{x^{n+1}}\odv{{x^\prime}^0}{{x^\prime}^{n+1}}=\pdv{{x^\prime}^0}{x^0}\odv{x^0}{x^{n+1}}+\sum_{k=1}^n \pdv{{x^\prime}^0}{x^k}\odv{x^k}{x^{n+1}}+\pdv{{x^\prime}^0}{x^{n+1}}
        \end{equation}
Taking into account that $x^0$ and $x^{n+1}$ are identified with the action and time variables respectively, we conclude that eq.~\eqref{transformationlaw2}
coincides with the transformation rule~\eqref{corresp}.

\end{section}
 \begin{section}{Symmetries, Noether's theorem and Killing vectors}\label{sec:sym}

The famous Noether's theorem relates the symmetries to conservation laws.
In the framework of Eisenhart formalism the trajectories are lifted to geodesics in $(n+2)$-dimensional (pseudo-)Riemannian manifold. Brinkmann metric is determined by
the initial Lagrangian; therefore, the symmetries of the latter can be lifted to the symmetries of the metric which, in turn, are determined by the relevant Killing vectors. Actually, since we are
considering null geodesics, it is sufficient to restrict ourselves to conformal Killing vectors,~i.e.~we are really interested in the classes of conformally equivalent metrics.
One can single out uniquely one metric from each class by choosing the normalization condition $g_{uw}=-1$.   
\newpage
Let us now extend the reasoning sketched above to the case of action-dependent Lagrangians. Assume that the transformations
        \begin{equation}
            x^i \mapsto x^i + \delta x^i(\underline{x},u,w)
        \end{equation}

        \begin{equation}
            u \mapsto u+\delta u(\underline{x},u,w)
        \end{equation}

        \begin{equation}
            w \mapsto w+\delta w(\underline{x},u,w)
        \end{equation}

define the symmetry of the Lagrangian~\eqref{afterreduction}.
The symmetry condition~\eqref{hergsym} reads here
        \begin{align}\label{preidentity}
            &\left(\frac{1}{2}\pdif{k}{h_{ij}}\odv{x^i}{u}\odv{x^j}{u}+\pdif{k}{A_i}\odv{x^i}{u}-\pdif{k}{V}\right)\delta x^k + \nonumber\\
            +\;&\left(h_{ki}\odv{x^i}{u}+A_k\right)\left(\odv{\delta x^k}{u}-\odv{x^k}{u}\odv{\delta u}{u}\right)+\nonumber\\
            +\:&\left(\frac{1}{2}\pdif{u}{h_{ij}}\odv{x^i}{u}\odv{x^j}{u}+\pdif{u}{A_i}\odv{x^i}{u}-\pdif{u}{V}\right)\delta u\;+ \nonumber\\
            +\;&\left(\frac{1}{2}\pdif{w}{h_{ij}}\odv{x^i}{u}\odv{x^j}{u}+\pdif{w}{A_i}\odv{x^i}{u}-\pdif{w}{V}\right)\delta w \nonumber\\
            +\;& \left(\frac{1}{2}h_{ij}\odv{x^i}{u}\odv{x^j}{u}+A_i\odv{x^i}{u}-V\right)\odv{\delta u}{u} - \odv{\delta w}{u}=0
        \end{align}
and
        \begin{equation}\label{1}
            \odv{x^k}{u}=\odv{x^l}{u}\,\pdif{l}{\delta x^k}+\pdif{u}{\delta x^k}+\odv{w}{u}\,\pdif{w}{\delta x^k}
        \end{equation}

        \begin{equation}
            \odv{\delta u}{u}=\odv{x^k}{u}\,\pdif{k}{\delta u}+\pdif{u}{\delta u}+\odv{w}{u}\,\pdif{w}{\delta u}
        \end{equation}

        \begin{equation}
            \odv{\delta w}{u}=\odv{x^k}{u}\,\pdif{k}{\delta w}+\pdif{u}{\delta w}+\odv{w}{u}\,\pdif{w}{\delta w}
        \end{equation}

        \begin{equation}\label{4}
            \odv{w}{u}=\frac{1}{2}h_{ij}\odv{x^i}{u}\odv{x^j}{u}+A_i \odv{x^i}{u} - V
        \end{equation}
Eqs.~\eqref{1}-\eqref{4}, when inserted into eq.~\eqref{preidentity} lead to the identity with the left-hand side being fourth degree polynomial in velocities $\odv{x^i}{u}$.
Therefore, one obtains five equations for the coefficients in front of consecutive powers of velocities. Fourth degree term yields
        \begin{equation}
            \left(h_{ij} h_{kl} + h_{ik}h_{jl}+h_{il}h_{jk}\right)\pdif{w}{\delta u}=0
        \end{equation}
and, after multiplying by $h^{ij}h^{kl}$,
        \begin{equation}\label{4deg}
            \pdif{w}{\delta u}=0
        \end{equation}
Third degree term leads to the relation
        \begin{align}\label{3deg}
            &\left(h_{il}h_{jk}+h_{jl}h_{ik}+h_{ij}h_{kl}\right)\pdif{w}{\delta x^l}+\nonumber\\
            -\;&\left(h_{ik}\pdif{j}{}+h_{ij}\pdif{k}{}+h_{jk}\pdif{i}{}\right)\delta u=0
        \end{align}
where~\eqref{4deg} has been already used.
By multiplying eq.~\eqref{3deg} by $h^{jk}$ one finds
        \begin{equation}\label{iden3}
            h_{il}\pdif{w}{\delta x^l}=\pdif{i}{\delta u}
        \end{equation}
Putting back~\eqref{iden3} into~\eqref{3deg} one finds that it is satisfied identically.

The remaining identities, corresponding to the terms of second, first and zero degree, read, respectively
        \begin{align}
            & \left(\delta u\,\pdif{u}{}-\pdif{u}{\delta u}\right)h_{ij} + \left(\delta w\,\pdif{w}{}-\pdif{w}{\delta w}\right)h_{ij}+A_k\,\pdif{w}{\delta x^k}\,h_{ij}+\nonumber\\
            +\;&\left(\pdif{i}{\delta u\, A_j}+\pdif{j}{\delta u}\, A_i\right) + \left(h_{ik}\,\pdif{j}{}+h_{jk}\,\pdif{i}{}\right)\delta x^k + \delta x^k \pdif{k}{h_{ij}}=0
        \end{align}

        \begin{align}
            &\left(\delta w\, \pdif{w}{}-\pdif{w}{\delta w}\right)A_i + \delta u\,\pdif{u}{A_i}-2V\,\pdif{i}{\delta u}-\pdif{i}{\delta w}+\nonumber\\
            +\;&\left(\pdif{k}{A_i}+A_k\,\pdif{i}{}\right)\delta x^k + h_{ik}\pdif{u}{\delta x^k}+A_i A_k \pdif{w}{\delta x^k}=0
        \end{align}

        \begin{align}\label{lastiden}
            -\delta x^k\, \pdif{k}{V}-\delta u\, \pdif{u}{V}-\pdif{u}{\delta u}\, V - \delta w\, \pdif{w}{V} - \pdif{u}{\delta w}+ V\,\pdif{w}{\delta w}+A_k\, \pdif{u}{\delta x^k}-A_k\, \pdif{w}{\delta x^k}\, V =0
        \end{align}
As in the standard case one can now find the Killing vector corresponding to the symmetries under consideration.
Namely, we put
        \begin{align}\label{killing}
            K\equiv K^\mu \pdif{\mu}{}&=K^i \pdif{i}{}+K^u\pdif{u}{}+K^w\pdif{w}{}=\nonumber\\
                                    &= \delta x^i\,\pdif{i}{} + \delta u\,\pdif{u}{} + \delta w\,\pdif{w}{}
        \end{align}
Using the identities~\eqref{4deg},~\eqref{iden3}-\eqref{lastiden} one finds that $K$ is a conformal vector field:
        \begin{equation}
            \nabla_\mu K_\nu + \nabla_\nu K_\mu =\lambda g_{\mu\nu}
        \end{equation}
with
        \begin{equation}
            \lambda\equiv \pdif{u}{\delta u}+\pdif{w}{\delta w}-A_i\,\pdif{w}{\delta x^i}
        \end{equation}
The conserved quantity related to the Killing field $K$ reads
        \begin{equation}\label{confint}
            K_\mu \odv{x^\mu}{\sigma}=g_{\mu\nu}K^\mu\odv{x^\nu}{\sigma}=\odv{u}{\sigma}Q
        \end{equation}

        \begin{equation}
            Q\equiv \bigg(h_{ij}\odv{x^j}{u}+A_i\bigg)\delta x^i - \left(\frac{1}{2}h_{ij}\odv{x^i}{u} \odv{x^j}{u}+V\right)\delta u - \delta w
        \end{equation}
where we have used eqs.~\eqref{metricgeneral},~\eqref{afterreduction} and~\eqref{killing}.
In the standard case $Q$ is the Noether charge ($\delta w$ accounts for total time derivative which, in general, enters Noether identity).
Due to eq.~\eqref{uddot} $\odv{u}{\sigma}$ is a constant.
However, in the general case, eq.~\eqref{ugeneral}, only the product $\odv{u}{\sigma}Q$ is a constant of motion.
The $u$ variable, which plays the role of time in reduced dynamics is no longer an affine parameter.
Given any parametrization of a geodesic one can find the affine parameter by solving an ordinary differential equation which, however, depends on the choice of particular geodesic.
In this sense the charge~\eqref{confint} is nonlocal.
Specifically, one has
        \begin{equation}
            \odv*{\left(K_\mu \odv{x^\mu}{\sigma}\right)}{\sigma}=\odv[order=2]{u}{\sigma}Q+\odv{u}{\sigma}\odv{Q}{\sigma}
        \end{equation}
Using~\eqref{uddotu2} we obtain
        \begin{align}\label{confintroc}
            \odv*{\left(K_\mu \odv{x^\mu}{\sigma}\right)}{\sigma}   &=-{\left(\odv{u}{\sigma}\right)}^{\!2} \pdv{\mathcal{L}}{w} Q + \odv{u}{\sigma}\odv{Q}{\sigma}=\nonumber\\
                                                                    &={\left(\odv{u}{\sigma}\right)}^{\!2}\left(-\pdv{\mathcal{L}}{w}Q+\odv{Q}{u}\right)=\nonumber\\
                                                                    &={\left(\odv{u}{\sigma}\right)}^{\!2} e^{\,\int^u\!\pdv{\mathcal{L}}{w}\odif{u^\prime}} \odv*{\left(e^{-\int^u\!\pdv{\mathcal{L}}{w}\odif{u^\prime}} Q\right)}{u}
        \end{align}
in agreement with the generalized Noether theorem (cf.~Appendix, eq.~\eqref{charge})

{Let us dwell on this point for a moment. For $w$-independent metric (i.e. action-independent Lagrangian) $\pdv{}{w}$ is the Killing vector field resulting in Noether symmetry
and the corresponding conservation law $p_w=\text{const.}$, following from eq.~\eqref{uddot}.
There are standard methods for reducing variables of Hamiltonian dynamics in the presence of conserved quantities (see, for example \cite{Arnold}).
However, in the presence of reparametrization invariance we believe a slightly different point of view is more fruitful.
Namely, as it is clearly seen from Section~\ref{sec:lift}, the existence of integral of motion allows us to ascribe to one of the two extra variables, $u$, the meaning of time (while $w$ is action).
The role of conservation law is that the relation between the evolution parameter $\sigma$ and time $u$, $u=-p_w \sigma$, is local, in the sense that it depends only on the values of dynamical variables at one point of trajectory.
The advantage of this way of thinking is that it retains its value in the general action-dependent case.
Then, in general there exists no (conformal) Killing field generalizing $\pdv{}{w}$ while one still expects to be able to replace $\sigma$ by $u$ in the role of evolution parameter.
While the absence of conservation law does not allow to look for simple local relation between $\sigma$ and $u$, we can use the generalization of eq.~\eqref{uddot}, eq.~\eqref{uddotu2}, which can be rewritten as
\begin{equation}
    \odif{\sigma}=e^{\int^u \pdif{w}\mathcal{L}\odif{u^\prime}}\odif{u}
\end{equation} 
The important difference is that the relation between $\sigma$ and $u$ is nonlocal, i.e. depends on particular trajectory.
In this way the local integrals of motion on the upper level become nonlocal on the reduced level (cf. eqs.~\eqref{confint}~-~\eqref{confintroc}).
In the action-dependent case the relation between initial and reduced dynamics goes beyond the standard framework. In the standard case the conserved charge, being a function of the actual values of dynamical variables,
makes the integration of the equations of motion easier (typically by lowering their order).
On the contrary, the nonlocal charge~\eqref{confintroc} is the functional of the whole previous dynamical history of the system so its utility as a tool for solving the equations of motion is limited.
However, all is not lost. Note that the nonlocal, exponential factor is universal.
Therefore, if the system enjoys two independent continuous symmetries, the ratio of the corresponding nonlocal charges is a local conserved charge which can be used for solving dynamical equations.
From the point of view of the ``large'' system the emergence of nonlocality appears to be quite natural.
The resulting nonlocal charge is obtained from the local one by changing the evolution parameter from $\sigma$ to $u$;
however, the latter is, in general, only possible provided the trajectory $u=u(\sigma)$ is known. 
The nonlocality can be also understood from the physical point of view.
It is well known that the Herglotz formalism is particularly useful in the description of dissipative systems.
Consider the energy conservation for such a system.
The change of kinetic energy equals the work performed by, say, friction; the latter is, obviously, a functional of trajectory.}
\end{section}
\begin{section}{Time versus action dependence}\label{sec:timeaction}

It is well known that the dissipative systems, demanding time-dependent Lagrangians for their description, can be also described with the help of action-dependent Lagrangians~\cite{TimeDependent}.
As a simple example consider the Lagrangian
        \begin{equation}
            L=e^{\gamma t}\left(\frac{1}{2} \left(\odv{\vec{x}}{t}\right)^{\!2}-V(\vec{x})\right)
        \end{equation}
Lagrange equations of motion read:
        \begin{equation}\label{eom}
            \odv[order=2]{\vec{x}}{t} + \gamma \odv{\vec{x}}{t} + \pdv{V}{\vec{x}}=0
        \end{equation}
Equivalently, eq.~\eqref{eom} can be derived from the following Lagrangian
        \begin{equation}
            \tilde{L}=\frac{1}{2}\left(\odv{\tilde{\vec{x}}}{\tilde{t}}\right)^{\!2}-V\big(\tilde{\vec{x}}\big)-\gamma\tilde{S}
        \end{equation}
Solving $\odv{S}{t}=L$ with the initial condition $S(t_0)=0$ one finds
        \begin{equation}\label{stimedep}
            S(t)=\int_{t_0}^t e^{\gamma t^\prime}\left(\frac{1}{2}\left(\odv{\vec{x}}{t}\right)^{\!2}(t^\prime)-V(\vec{x}(t^\prime))\right)\odif{t^\prime}
        \end{equation}
Analogously, solving $\odv{\tilde{S}}{\tilde{t}}=\tilde{L}$ with the same initial conditions yields
        \begin{equation}\label{sactiondep}
            \tilde{S}(\tilde{t})=e^{-\gamma \tilde{t}}\int_{\tilde{t}_0}^{\tilde{t}} e^{\gamma t^\prime}\left(\frac{1}{2}\left(\odv{\tilde{\vec{x}}}{\tilde{t}}\right)^{\! 2}\!\left(t^\prime\right)-V(\tilde{\vec{x}}(t^\prime))\right)\odif{t^\prime}
        \end{equation}
\pagebreak
        Comparing eqs.~\eqref{stimedep}~and~\eqref{sactiondep} we infer that the above descriptions are related by transformation
        \begin{equation}\label{transx}
            \tilde{\vec{x}}=\vec{x}
        \end{equation}

        \begin{equation}\label{transt}
            \tilde{t}=t
        \end{equation}

        \begin{equation}\label{transs}
            \tilde{S}=e^{-\gamma t}S
        \end{equation}
Indeed, one finds from~\eqref{corresp}
        \begin{equation}
            \tilde{L}=e^{-\gamma t} L-\gamma \tilde{S}
        \end{equation}
On the level of Brinkmann metrics one has
        \begin{equation}
            \odif{s}^2 = e^{\gamma u} {\odif{\vec{x}}}^{\,2} - 2 e^{\gamma u} V \odif{u}^{2} - 2\odif{u,w}
        \end{equation}

        \begin{equation}
            \odif{\tilde{s}}^2 = \odif{\tilde{\vec{x}}}^{\,2} - 2(V+\gamma \tilde{w})\odif{\tilde{u}}^2 - 2\odif{\tilde{u},\tilde{w}}
        \end{equation}

Eqs.~\eqref{transx}-\eqref{transs} suggest the following transformation:
        \begin{equation}
            \tilde{\vec{x}}=\vec{x}
        \end{equation}

        \begin{equation}
            \tilde{u}=u
        \end{equation}

        \begin{equation}
            \tilde{w}=e^{-\gamma u} w
        \end{equation}
which implies the following relation
        \begin{equation}
            \odif{\tilde{s}}^2=e^{-\gamma \tilde{u}} \odif{s}^2
        \end{equation}
The metrics $\odif{s}^2$ and $\odif{\tilde{s}}^2$ are conformally equivalent so they share the same set of null geodesics.
    \end{section}

 \section{Scaling symmetry}\label{sec:scaling}
Another nice example of the application of Herglotz formalism is the scaling symmetry {(cf. also \cite{Sloan})}.

Let $L(q,\odv{q}{t},t)$ be a standard (i.e.~action-independent) Lagrangian.
Consider the following generalized point transformations:
\begin{align}\label{generalizedpoint}
    q^\prime&=q^\prime(q,t)\nonumber\\
    t^\prime&=t^\prime(q,t)\\
    S^\prime&=\Lambda S + f(q,t),\quad \Lambda \in \mathbb{R}\nonumber
\end{align}

Then the symmetry condition~\eqref{hergsym} takes the form:
\begin{equation}\label{conditionscaling}
    \mathcal{L}\left(q^\prime,\odv{q^\prime}{t^\prime},t^\prime\right)\odv{t^\prime}{t}=\Lambda \mathcal{L}\left(q,\odv{q}{t},t\right)+\odv{f\left(q,t\right)}{t}
\end{equation}
which coincides with eq. (II.1) from~\cite{Scaling}. The infinitesimal form of eq.~\eqref{generalizedpoint} reads:
\begin{align}
    t^\prime=t+\delta t(q,t)\nonumber\\
q^\prime = q + \delta q(q,t)\\
S^\prime=S +\delta \Lambda S + \delta f(q,t)\nonumber
\end{align}
and the Noether identity~\eqref{charge} reads ($\pdv{\mathcal{L}}{S}=0$!)
\begin{equation}
    \odv*{\left(\pdv{\mathcal{L}}{\dot{q}}\delta q - H \delta t - \delta f - \delta \Lambda S\right)}{t}=0
\end{equation}
Now, taking into account that $S$ is the standard action, $S=\int^t L \odif{\tau}$, we arrive at the eq. (II.4) from~\cite{Scaling}.
\newpage

Consider now the Eisenhart lift. Assume that the Lagrangian~\eqref{afterreduction} is invariant under the transformation:
\begin{align}\label{brinkmannscaling}
    {x^\prime}^i&={x^\prime}^i\left(\underline{x},u\right)\\
    u^\prime &= u^\prime\left(u\right)\nonumber 
\end{align}
in the sense that eq.~\eqref{conditionscaling} is obeyed (note that, due to the quadratic dependence of $\mathcal{L}$ on velocities, this can happen only provided $u^\prime$ is a function of $u$ only):
\begin{align}
    &\left(\frac{1}{2}h_{ij}(\underline{x}^\prime,u^\prime) \odv{{x^\prime}^i}{u^\prime} \odv{{x^\prime}^j}{u^\prime} + A_i(\underline{x}^\prime,u^\prime)\odv{{x^\prime}^i}{u^\prime}-V(\underline{x}^\prime,u^\prime)\right)\odv{u^\prime}{u}=\nonumber\\
    =\Lambda &\left(\frac{1}{2}h_{ij}(\underline{x},u) \odv{x^i}{u} \odv{x^j}{u} + A_i(\underline{x},u)\odv{x^i}{u}-V(\underline{x},u)\right)+\odv{f}{u}\left(\underline{x},u\right)
\end{align}

which can be rewritten as:
\begin{align}\label{rewritten}
     &\frac{1}{2}h_{ij}(\underline{x}^\prime,u^\prime) \odif{{x^\prime}^i,{x^\prime}^j} + A_i(\underline{x}^\prime,u^\prime)\odif{{x^\prime}^i,u^\prime}-V(\underline{x}^\prime,u^\prime)\odif{u^\prime}^2=\nonumber\\
    =\Lambda \odv{u^\prime}{u} &\left(\frac{1}{2}h_{ij}(\underline{x},u) \odif{x^i,x^j} + A_i(\underline{x},u)\odif{x^i,u}-V(\underline{x},u)\odif{u}^2\right)+\odv{u^\prime}{u}\odif{u,f}
\end{align}

Let us supply~\eqref{brinkmannscaling} with the transformation rule:
\begin{equation}\label{wtransformation}
    w^\prime = \Lambda w + f(\underline{x},u)
\end{equation}

so that
\begin{equation}\label{dw}
    \odif{w^\prime}=\Lambda \odif{w} + \odif{f}
\end{equation}

and 

\begin{equation}\label{dwdu}
    \odif{u^\prime, w^\prime}=\odv{u^\prime}{u}\Lambda \odif{u,w}+\odv{u^\prime}{u}\odif{u,f}
\end{equation}

It follows then from eqs.~\eqref{metricgeneral},~\eqref{rewritten}~and~\eqref{dwdu} that, under the transformations~\eqref{brinkmannscaling},~\eqref{wtransformation}
the metric transforms conformally:
\begin{equation}
    (\odif{s^\prime})^2=\Lambda \odv{u^\prime}{u}\odif{s}^2
\end{equation} 

{An interesting class of examples, also related to scaling symmetry, where Eisenhart lift provides a nice picture of relevant dynamics are homogeneous cosmologies.
Consider, for example, the metric
\begin{equation}\label{metricbianchi}
    \odif{s}^2=\rho\left(-\odif{t}^2+e^{-\frac{x}{\sqrt{3}}+y}\odif{x}^2+e^{-\frac{x}{\sqrt{3}}-y}\odif{y}^2+e^{\frac{2x}{\sqrt{3}}}\left(x\odif{y}+\odif{z}\right)^2\right)
\end{equation}
which belongs to Bianchi class A models.
When inserted into \mbox{Einstein-Hilbert} Lagrangian this metric produces the effective Lagrangian.
It has been shown~\cite{Sloan} that due to the scaling properties of the metric the space-time dynamics can be described by the action-dependent Lagrangian
\begin{equation}\label{SloanLag}
    L=\frac{1}{2}\left(\dot{x}^2+\dot{y}^2\right)-\frac{1}{2}e^{\frac{4x}{\sqrt{3}}}+\frac{S^2}{6}
\end{equation}
with dots denoting time derivatives.
Once the Lagrange-Herglotz equations following from \eqref{SloanLag} are solved, the scaling factor $\rho$ is recovered from
\begin{equation}
    \frac{\dot{\rho}}{\rho}=-\frac{S}{3}
\end{equation}
According to our discussion the relevant Eisenhart metric reads
\begin{equation}\label{SloanMetric}
    \odif{s_E}^2=\odif{x}^2+\odif{y}^2-\left(e^{\frac{4x}{\sqrt{3}}}-\frac{w^2}{3}\right)\odif{u}^2-2\odif{u,w}
\end{equation}
}

{
We see that the dynamics of homogeneous Universe can be equivalently described as geodesic motion in the metric \eqref{SloanMetric}.
It should be noticed that by inserting directly the metric \eqref{metricbianchi} into Einstein-Hilbert Lagrangian one obtains a standard Lagrangian involving $x,y$ and $\rho$.
Therefore, the Eisenhart lift can be applied at this stage. However, the resulting metric is singular.}
 \begin{section}{Summary}\label{sec:summary}
    
We have shown that the idea of Eisenhart lift can be extended to the case of general Brinkmann metric with components depending
on all coordinates. The price to be paid is that one has to start with Herglotz's formalism which is more general than the Lagrangian one.
So, starting with an action-dependent Lagrangian and performing essentially the same steps as in the case of standard Eisenhart lift one arrives at the most general form of Brinkmann metric.
The important difference is that the coordinate $u$, identified with the physical time, cannot any longer serve as an affine parameter for null geodesics in Brinkmann metric (cf.~eqs.~\eqref{uddot}~and~\eqref{ugeneral}).
This has profound consequences. The conserved charge, eq.~\eqref{confint}, involves the derivative of $u$ with respect to an affine parameter which, when expressed in terms of $u$, results in nonlocal expression for the conserved quantity,
in agreement with known results~\cite{2NoetherTheorem,1NoetherTheorem,NoetherActionDep}.

We have also shown on a simple example that the equivalence between time-dependent and action-dependent descriptions of dynamics translates into the conformal equivalence of relevant Brinkmann metrics.
It would be interesting to find the Hamiltonian description of the above reduction procedure. This is important because the Noether theorem can be extended to Hamiltonian formalism: the conserved quantities result also from symmetries described by
canonical transformations which do not reduce to the point ones. On the level of lifted dynamics this should correspond to the existence of (conformal) Killing tensors.
   
\end{section}
   \newpage
    \appendix
    
    \counterwithin*{equation}{section}
    \renewcommand\theequation{\thesection\arabic{equation}}
    
\begin{section}{Herglotz variational principle}

{Let us start with the standard Lagrangian formalism. Given a Lagrange function
\begin{equation}\label{lagnos}
L=L(q,\dot{q},t)
\end{equation}
(for simplicity we consider one degree of freedom; the generalization to any finite number of degrees of freedom is straightforward)
we consider the set of (sufficiently differentiable) functions $q=q(t)$, defined on the interval $\left[t_0,t_1\right]$ and obeying $q(t_0)=q_0$, $q(t_1)=q_1$
with $q_0, q_1$ fixed.
Choosing an arbitrary fixed constant $S(t_0)$ we define
\begin{equation}\label{actiont1}
    S(t_1)=\int_{t_0}^{t_1} L(q(t),\dot q(t),t)\odif{t} + S(t_0)
\end{equation}
Then $S$ becomes a functional of $q(t)$:
\begin{equation}\label{sfunctional}
    S(t_1)\equiv S\left[t_1;q(t)\right]
\end{equation}
According to the standard variational principle the physical trajectory is the one for which $S(t_1)$ is stationary under infinitesimal variations
\begin{equation}
    q(t)\mapsto q(t)+\delta q(t),\quad \delta q(t_0)=0=\delta q(t_1)
\end{equation}
in the class of functions we are considering:
\begin{equation}
    \delta S(t_1)=0
\end{equation}
This can be reformulated as follows: given any function $q(t)$, defined on~$\left[t_0,t_1\right]$, one considers the differential equation
\begin{equation}\label{odes}
    \odv{S(t)}{t}=L(q(t),\dot{q}(t),t)
\end{equation}
on the interval $\left[t_0,t_1\right]$ with the initial condition $S(t=t_0)=S(t_0)$.
Once~\eqref{odes} is solved we consider $S(t_1)$ as the functional of $q(t)$ and look for a particular trajectory $q=q(t)$ at which $S(t_1)$ is stationary.
Due to the fact that the right-hand side of~\eqref{odes} depends only on $t$ (explicitly, once $q(t)$ is given), eqs.~\eqref{actiont1} and~\eqref{odes} are obviously equivalent.}

{
The above suggests the following straightforward generalization. Assume $L$ depends also on $S$,
\begin{equation}
    L=L(q,\dot{q},t,S)
\end{equation}
and consider again the set of sufficiently differentiable functions $q=q(t)$ on~$\left[t_0,t_1\right]$.
Once $q(t)$ is given,
\begin{equation}\label{odvss}
    \odv{S}{t}=L(q(t),\dot{q}(t),t,S)
\end{equation}
\begin{equation}\label{ivp}
    S(t=t_0)=S(t_0)
\end{equation}
becomes ordinary first order differential equation for $S=S(t)$ with an initial condition \eqref{ivp}.
Once it is solved one obtains the value $S(t_1)$ as a functional of $q(t)$
\begin{equation}
    S(t_1)\equiv S\left[t_1;q(t)\right]
\end{equation} 
Again, the physical trajectory can be defined by demanding stationarity of~$S(t_1)$ as the functional of $q(t)$, $t_0\leq t \leq t_1$. Obviously, if $L$ does not depend on $S$, one reproduces the standard formalism.}

Herglotz formalism shares many properties of the Lagrangian one. Euler-Lagrange equations are replaced by eq.~\eqref{odvss} together with
        \begin{equation}
            \pdv{L}{q}-\odv*{\left(\pdv{L}{\dot{q}}\right)}{t}+\pdv{L}{S}\pdv{L}{\dot{q}}=0
        \end{equation}
\newpage
The generalized point transformations

        \begin{equation}\label{Attransf}
            t^\prime=t^\prime(t,q,S)
        \end{equation}

        \begin{equation}
            q^\prime=q^\prime(t,q,S)
        \end{equation}

        \begin{equation}\label{Astransf}
            S^\prime=S^\prime(t,q,S)
        \end{equation}

preserve the form of the equations of motion provided the Lagrangian transforms according to the formula:
        
        \begin{equation}\label{corresp}
            L^\prime \odv{t^\prime}{t}=\pdv{S^\prime}{S}L+\pdv{S^\prime}{q}\odv{q}{t}+\pdv{S^\prime}{t}
        \end{equation}

        Point transformations~\eqref{Attransf}-\eqref{Astransf} define a symmetry if the new Lagrangian has the same functional form, i.e.

        \begin{equation}\label{hergsym}
            L\left(q^\prime,\odv{q^\prime}{t^\prime},t^\prime,S^\prime\right)\odv{t^\prime}{t}=\pdv{S^\prime}{S}L\left(q,\odv{q}{t},t,S\right)+\pdv{S^\prime}{q}\odv{q}{t}+\pdv{S^\prime}{t}
        \end{equation}

Continuous symmetry transformations generate, via generalized Noether's theorem, the corresponding conservation laws.
Consider namely the infinitesimal version of~\eqref{Attransf}-\eqref{Astransf}:
        
        \begin{equation}
            t^\prime=t+\delta t(t,q,S)
        \end{equation}

        \begin{equation}
            q^\prime=q+\delta q(t,q,S)
        \end{equation}

        \begin{equation}
            S^\prime=S+\delta S(t,q,S);
        \end{equation}

then, for physical trajectories, the following relation is fulfilled
        
        \begin{equation}\label{charge}
            \odv*{\left(e^{-\int^t\!\pdv{L}{S}\odif{t^\prime}}\left(L\delta t+\pdv{L}{\dot{q}}\left(\delta q-\dot{q}\delta t\right)-\delta S\right)\right)}{t}=0
        \end{equation}

Eq.~\eqref{charge} reduces to the standard conservation law provided $\pdv{L}{S}=0$. In general, however,~\eqref{charge} is more complicated
since the exponential term depends on the whole past history of the system. 

Therefore, it ceases to be a useful tool for integrating the equations of motion.
It can be shown that it is also possible to generalize appropriately the Hamiltonian formalism, Poisson brackets, canonical transformations,~etc.~to~the~case of action dependent Lagrangians.
We will not dwell on it here.

\end{section}
\begin{section}{Reparametrization invariance}

The Lagrangian defined by eq.~\eqref{wdot} in the main text reads
        \begin{equation}\label{appendixltilde}
            \tilde{\mathcal{L}}=\frac{1}{2}h_{ij}\frac{\dot{x}^i\dot{x}^j}{\dot{u}} + A_i \dot{x}^i - V \dot{u}
        \end{equation}
We have shown that the final equations of motion can be written in terms of $u$-derivatives only. This can be also inferred from the homogeneity of $\tilde{\mathcal{L}}$ in velocities.
Indeed,~\eqref{appendixltilde} implies:
        \begin{equation}\label{identity}
            \dot{x}^k \pdv{\tilde{\mathcal{L}}}{\dot{x}^k}+\dot{u}\pdv{\tilde{\mathcal{L}}}{\dot{u}}=\tilde{\mathcal{L}}
        \end{equation}
The equations of motion read:
        \begin{equation}\label{eqxk}
            \pdv{\tilde{\mathcal{L}}}{x^k}-\odv*{\left(\pdv{\tilde{\mathcal{L}}}{\dot{x}^k}\right)}{\sigma}+\pdv{\tilde{\mathcal{L}}}{w}\pdv{\tilde{\mathcal{L}}}{\dot{x}^k}=0
        \end{equation}

        \begin{equation}\label{equ}
            \pdv{\tilde{\mathcal{L}}}{u}-\odv*{\left(\pdv{\tilde{\mathcal{L}}}{\dot{u}}\right)}{\sigma}+\pdv{\tilde{\mathcal{L}}}{w}\pdv{\tilde{\mathcal{L}}}{\dot{u}}=0
        \end{equation}
Taking into account that $\tilde{\mathcal{L}}=\tilde{\mathcal{L}}(\underline{x},u,w,\dot{\underline{x}},\dot{u})$ and $\dot{w}=\tilde{\mathcal{L}}$ one easily finds from~\eqref{identity}:
        \begin{equation}
            \dot{x}^k\eqref{eqxk}_k+\dot{u}\eqref{equ} \equiv 0
        \end{equation}
as an identity. Therefore,~\eqref{eqxk} implies~\eqref{equ}.
Now, $\tilde{\mathcal{L}}$ obeys:  
        \begin{equation}
            \tilde{\mathcal{L}}\left(\underline{x},u,w,\underline{\odv{x}{\sigma^\prime}},\odv{u}{\sigma^\prime}\right) = \odv{\sigma}{\sigma^\prime}\tilde{\mathcal{L}}\left(\underline{x},u,w,\underline{\odv{x}{\sigma}},\odv{u}{\sigma}\right)   
        \end{equation}
and it is straightforward to check that~\eqref{eqxk} is invariant under the reparametrization $\sigma \mapsto \sigma^\prime$.

\end{section}  \newpage

    \printbibliography
\end{document}